\documentclass[manuscript]{aastex}

\usepackage{rotating}
\usepackage{tikz}
\usepackage{amsmath}
\usepackage{gensymb}
\usepackage{booktabs}
\usepackage{pdflscape}

\slugcomment{The Astronomical Journal}

\shorttitle{Irregular Satellite Photometry}
\shortauthors{Graykowski et al.}

\begin{document}

\title{Colors and Shapes of the Irregular Planetary Satellites }


\author{Ariel Graykowski$^{1}$ and  David Jewitt$^{1,2}$
}

\affil{$^1$Department of Earth, Planetary and Space Sciences,
UCLA, 595 Charles Young Drive East, Los Angeles, CA 90095-1567\\
$^2$Department of Physics and Astronomy, University of California at Los Angeles, 430 Portola Plaza, Box 951547, Los Angeles, CA 90095-1547\\
}

\email{ajgraykowski@ucla.edu}

\begin{abstract}

It is widely recognized that the irregular satellites of the giant planets were captured from initially heliocentric orbits.  However, the mechanism of capture and the source region from which they were captured both remain unknown. We present an optical color survey of 43 irregular satellites of the outer planets conducted using the LRIS camera on the 10-meter telescope at the Keck Observatory in Hawaii. The measured colors are compared to other planetary bodies in search for similarities and differences that may reflect upon the origin of the satellites. We find that ultrared matter (with color index B-R $\ge$ 1.6), while abundant in the Kuiper belt and Centaur populations, is depleted from the irregular satellites. We also use repeated determinations of the absolute magnitudes to make a statistical estimate of the average shape of the irregular satellites. The data provide no evidence that the satellites and the main-belt asteroids are differently shaped, consistent with collisions as the major agent shaping both.

\end{abstract}


\section{Introduction}

Irregular satellites are broadly distinguished from regular satellites by their orbital characteristics. Regular satellites occupy nearly circular, low eccentricity orbits deep within the Hill spheres of their respective planets.  In contrast, the irregular satellites orbit at distances up to 0.5-0.6 Hill radii and are subject to significant torques from the Sun even while remaining bound to the host planets. The irregulars also have large eccentricities, $e$ = 0.1 to 0.7, and inclinations, $i$, many with $i >$ 90\degr~(Jewitt and Haghighipour 2007, Nicholson et al.~2008). Only  the  giant planets possess irregular satellites. The currently known  irregular satellite populations, as well as the numbers of  irregular satellites observed in this survey, are listed in Table \ref{number}.

The size distributions  of the irregular satellites  of each giant planet are similarly shallow, roughly consistent with differential power laws having index $q$ = -2   suggesting capture from  a common source region by a common mechanism (Sheppard and Jewitt 2003, Jewitt and Sheppard 2005). Three main classes of capture mechanism have been proposed. 1) Pull-down capture  relies on the runaway accretion phase of planetary growth, when the Hill radius of the planet grew rapidly (Heppenheimer and Porco, 1977). Nearby  bodies might have been permanently captured if the Hill radius expanded on a timescale short compared to the residence time within the Hill spheres. One argument against pull-down capture as a general mechanism is that the ice giants Uranus and Neptune have relatively little H and He in their gaseous envelopes, limiting the effects of runaway growth. 2) In gas-drag, the extended gaseous envelopes of the forming giant planets are supposed to frictionally dissipate the energy of  passing bodies, leading to permanent capture (Pollack et al.~1979).  This model relies on fine-tuning of the timing, because the collapse of the gaseous envelope is thought to have been rapid.  Capture by gas drag is again less attractive for Uranus and Neptune than for Jupiter and Saturn because the ice giants contain a much smaller fraction of their total mass in gas (Jewitt and Sheppard 2005; Jewitt and Haghighipour, 2007). 3) Accordingly, most recent work has focused on capture by three-body interactions, considered first by Colombo and Franklin (1971), since this mechanism is independent of the gas content and growth physics of the host planet.  In three body reactions, gravitational scattering between two bodies in the circumplanetary environment can, statistically, lead to the ejection of one and the capture of the other.   

In this paper we present measurements of the magnitudes and colors of 43 irregular satellites of the four giant planets taken using the Keck I 10 m telescope.  The new data are compared with published measurements of smaller samples (Grav et al.~2003, 2004, Grav and Bauer 2007, Rettig et al.~2001)

\section{Observations}

The data were collected over nine nights between  2008 March and  2015 December at the W. M. Keck Observatory on Mauna Kea, Hawaii using the LRIS instrument on the 10 meter Keck I telescope (Oke et al.~1995, Table \ref{geometry}). The data used were all taken under photometric conditions with the telescope tracked at non-sidereal rates to  follow the motion of each satellite. Most satellites were observed on multiple nights in order to check for repeatability of the measurements.  We used  the B, V, and R filters, for which the central wavelengths, $\lambda_c$, and full-widths at half maxima, FWHM, are B (4370\AA , 878\AA), V  (5473 \AA, 948 \AA) and R (6417 \AA, 1185 \AA). The images were flat-fielded using composites of images recorded from an illuminated patch inside the Keck dome and photometrically calibrated using observations of  stars with Sun-like colors from Landolt (1992).  

\begin{figure}
\epsscale{0.75}
\plotone{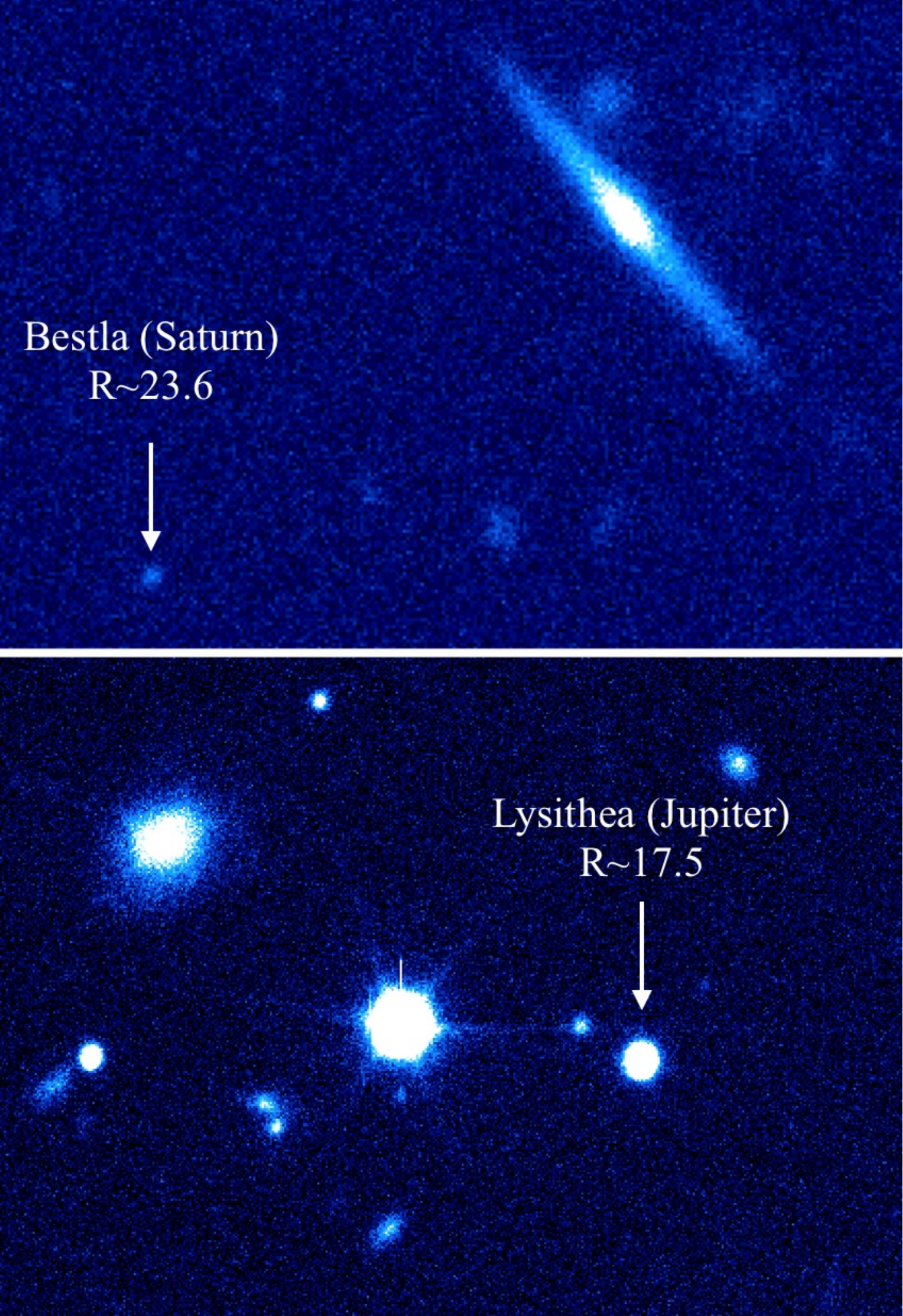}
\caption{Sample images of irregular satellites (top) Bestla, a faint irregular satellite of Saturn (bottom) Lysithea, a bright irregular satellite of Jupiter. \label{bestlalysithea}}
\end{figure}

Using IRAF, the images were reduced and aperture photometry was obtained using the APPHOT package. 
By trial and error, we used a photometry annulus with radius 1.35-2.03\arcsec ($\sim$1.5 x FWHM), depending on the seeing, and obtained an estimate of the sky background from a contiguous annulus  1.35\arcsec~wide. For very faint satellites, we used two-aperture photometry. With this method, we chose a small aperture based on the FWHM of the object, and used it to measure the targeted satellite as well as brighter field stars. Then we chose a larger aperture in order to measure the total flux from the selected field stars. We calculated the fraction of light that was left out of the measurement from the smaller aperture, and used it to correct the magnitude of the satellite to obtain its apparent magnitude. We observed satellites with apparent magnitude, R, as bright as $\sim$17.5 and as faint as $\sim$25.0 magnitude. To show the visual difference of this magnitude range, Figure (\ref{bestlalysithea}) compares an image of a faint ($\sim$23.6) and a bright ($\sim$17.5) satellite observed in this work.  

\section{Results}

The results of the photometry are listed in Table \ref{geometry} with $\pm$1$\sigma$ standard errors. Not all satellites were observed in all three filters (B, V, and R) and therefore 
 not all have equal numbers of color measurements.  In total, we measured  20 Jovian, 14 Saturnian, 6 Uranian, and 3 Neptunian satellites.  

The apparent magnitudes were  converted to absolute magnitudes, $H_V$, defined as the magnitude corrected to unit heliocentric and geocentric distance ($r_H$ and $\Delta$, respectively) and to phase angle  $\alpha$ = 0\degr.  For the apparent V magnitude, this correction is 

\begin{equation}
H_V = V - 5\log_{10}(r_H \Delta) - \beta \alpha
\label{abs}
\end{equation}

\noindent where $\beta$ is the phase function  representing the angular dependence of the scattered sunlight on $\alpha$.  For simplicity, we assumed $\beta$ = 0.04 magnitudes degree$^{-1}$, consistent with values measured in low albedo solar system objects (Tedesco and Baker 1981, Jewitt et al.~1998, Rettig et al.~2001). Equivalent relations were also used to compute the absolute B and R filter magnitudes.  

\begin{figure}
\epsscale{0.75}
\plotone{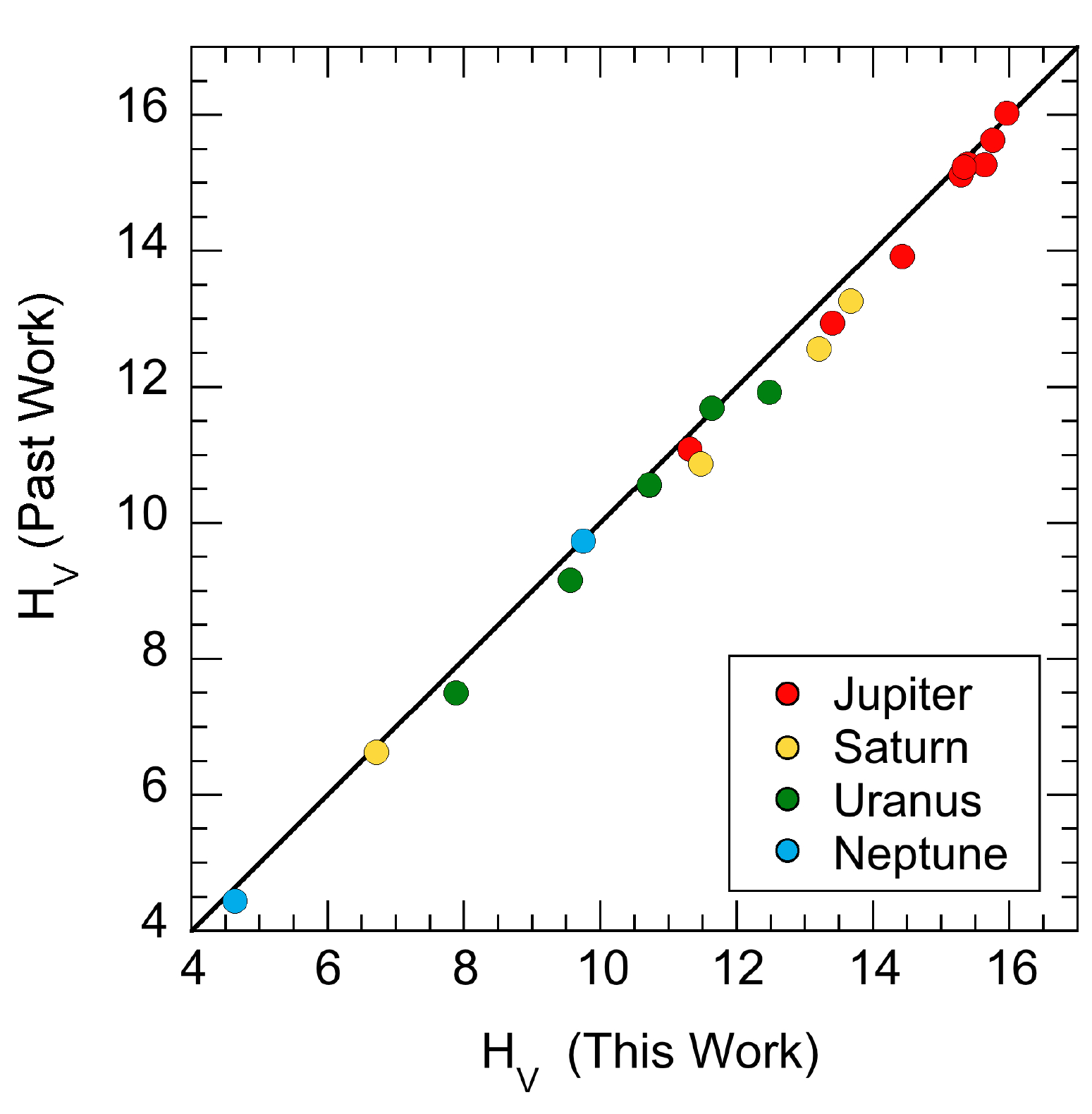}
\caption{ $H_V$ values from this survey plotted against $H_V$ values from Grav et al.~2003, 2004 and Grav and Bauer 2007. Only satellites measured in both surveys are plotted. The $H_V$ from other surveys are systematically brighter than in the present work. 
\label{hvcompare}}
\end{figure}

Figure (\ref{hvcompare}) compares $H_V$ magnitudes from this work with $H_V$ magnitudes from previous surveys by Grav et al.~(2003), Grav et al.~(2004) and Grav and Bauer (2007). The average error bars are on the order of 0.04 magnitudes, smaller than the data point symbols, and therefore do not appear in the figure. Measurements in perfect agreement should plot on the diagonal line in the figure. Some scatter about the line is expected because of measurement errors, and because each satellite possesses a rotational lightcurve, presenting a variable brightness to the observer. In fact, most satellites fall slightly below the diagonal line, indicating systematic differences between our measurements and those in the literature.   Possible reasons for these systematic offsets include slight differences in the filters employed, as well as differences in the way the phase function (Equation \ref{abs}) was treated. 

\begin{figure}
\epsscale{0.75}
\plotone{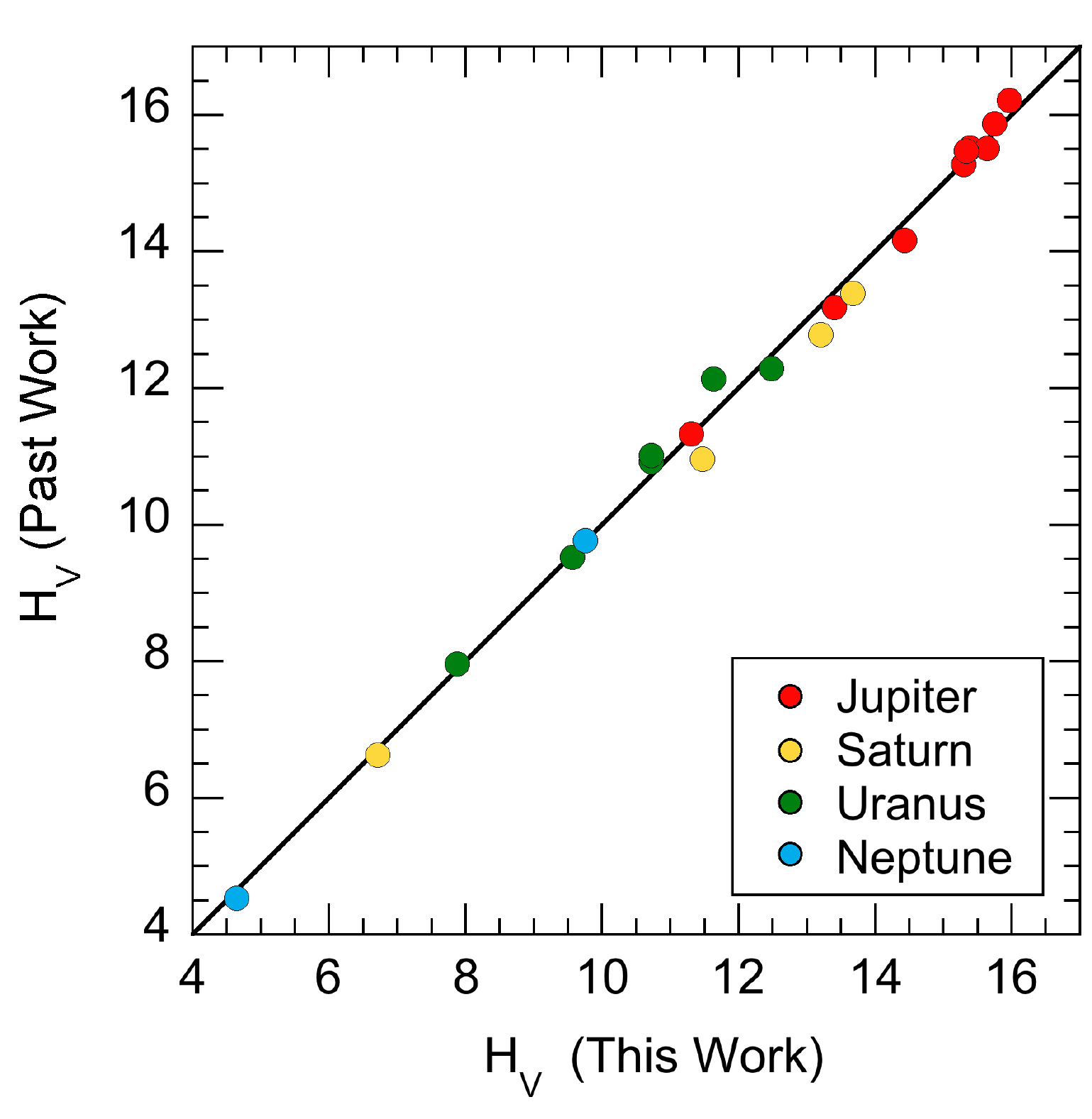}
\caption{ Same as Figure (\ref{hvcompare}) except the values from Grav et al.~2003, 2004 and Grav and Bauer 2007 now reflect $H_V$ values calculated with $\beta$ = 0.04 rather than $\beta$ = 0.38 or the Bowell et al.~(1989) phase function where G = 0.15 as used in the original work. The data no longer consistently fall below the line as in Figure (\ref{hvcompare}), showing that this systematic difference in $H_V$ values is a result of the choice of phase function. 
\label{hvcompareBETA}}
\end{figure}

The major uncertainty in the phase function correction lies in the treatment of the possible opposition surge. For example, Grav et al.~(2004) assumed $\beta$ = 0.38 magnitudes degree$^{-1}$ for the satellites of Uranus and Neptune in order to account for small-angle brightening.  Grav et al.~(2003, 2007) and Rettig et al.~(2001) instead used the Bowell et al.~(1989) phase function with parameter $G$ = 0.15, which provides for a more modest surge.  Bauer et al.~(2006) found that the magnitude of the opposition surge varies widely from satellite to satellite, meaning that we cannot adopt any universal value. To assess the impact of the various assumed phase functions, we recomputed the $H_V$ magnitudes from the photometry of Grav et al.~(2003, 2007)  and Rettig et al.~(2001) assuming  $\beta$ = 0.04 magnitudes degree$^{-1}$ for all objects, consistent with the value used in the analysis of the current data. Figure (\ref{hvcompareBETA}) shows that the systematic differences of Figure (\ref{hvcompare}) largely disappear, showing that the offsets result from phase and are not intrinsic to the data.  

\begin{figure}
\epsscale{0.75}
\plotone{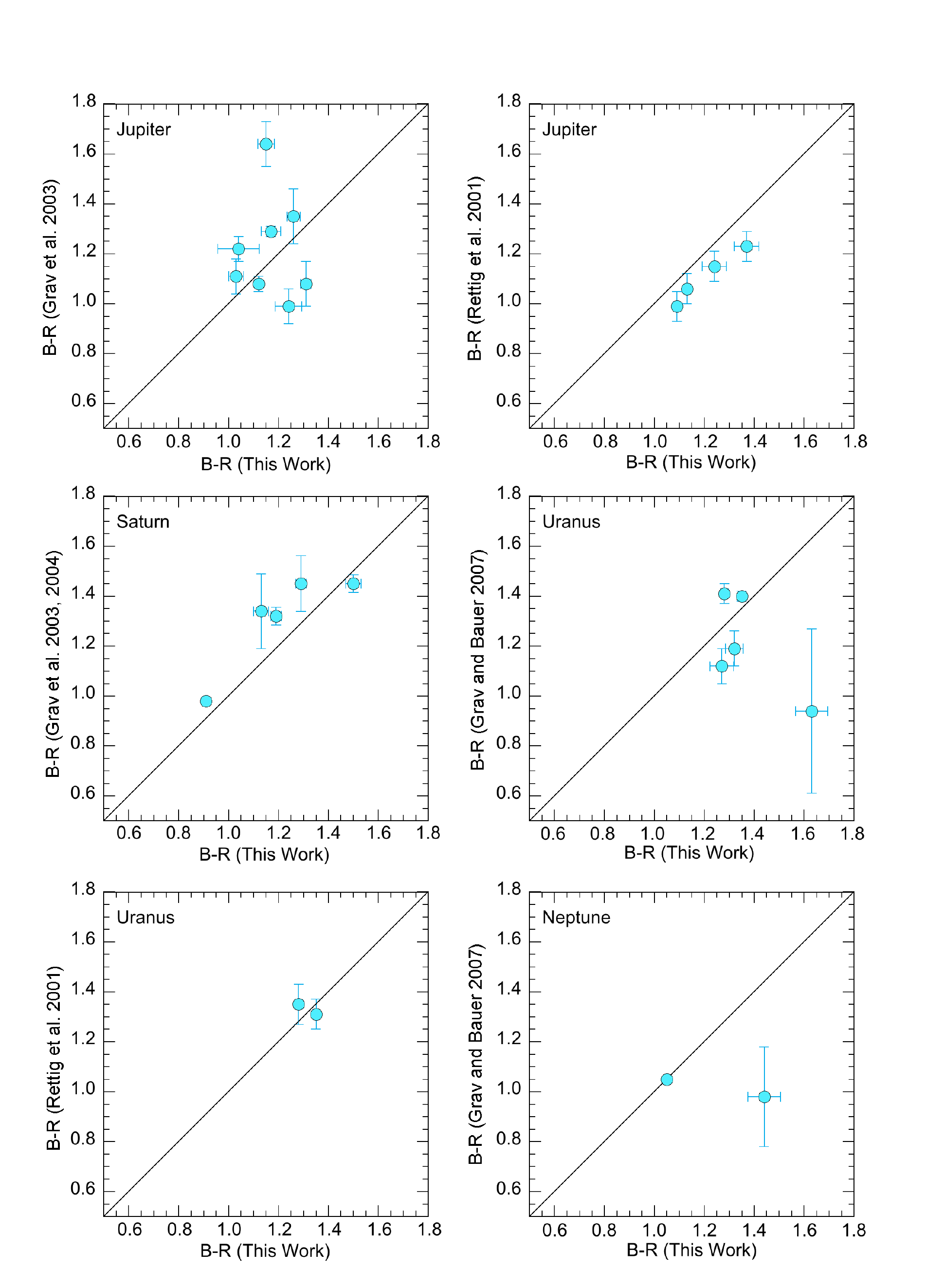}
\caption{The horizontal axis on each graph shows our B-R data and the vertical axis shows B-R data from previous surveys. The diagonal line shows where the measurements are equal.
\label{brbr}}
\end{figure}

\begin{figure}
\epsscale{0.9}
\plotone{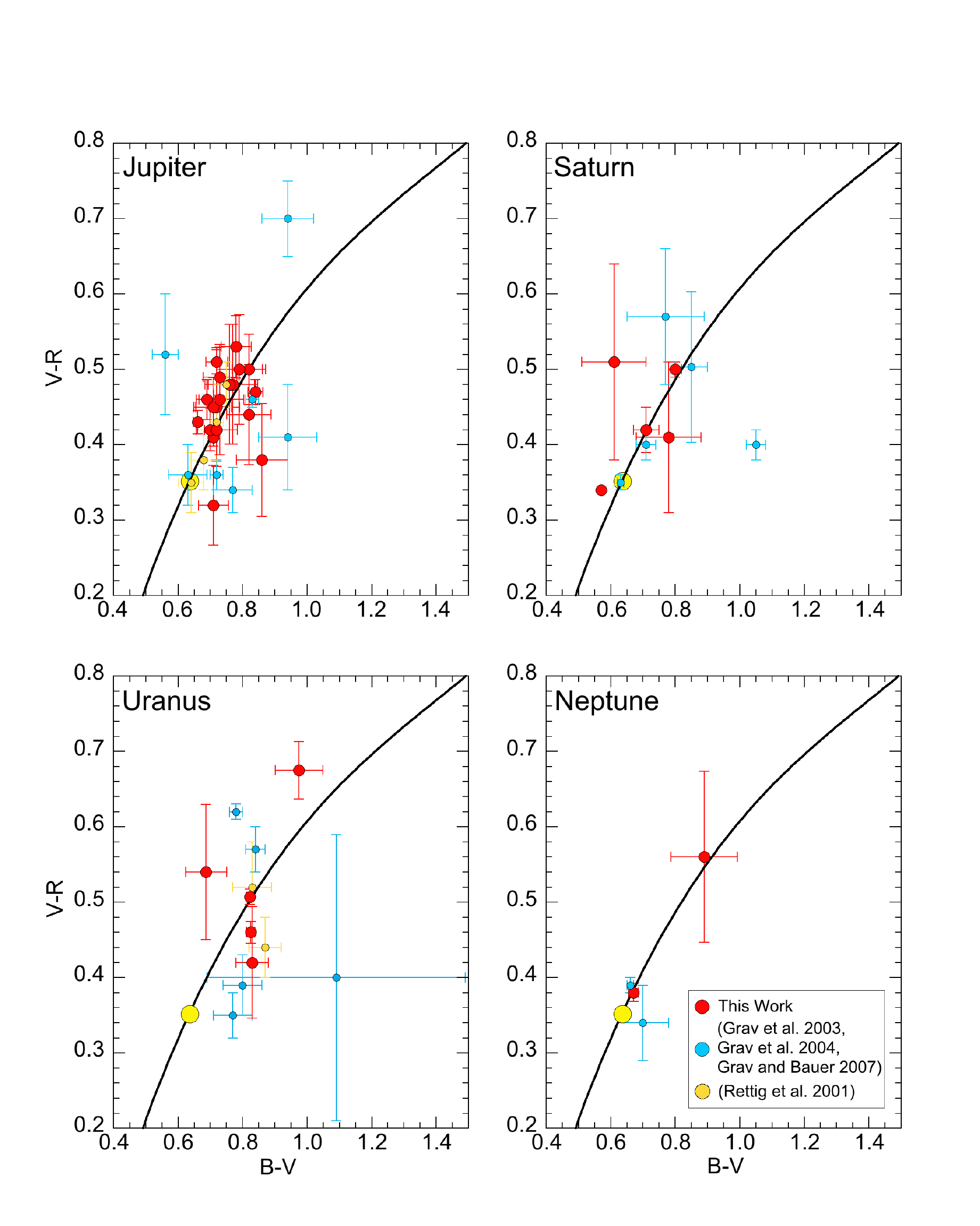}
\caption{Color vs. color plots of the irregular satellites at each of the giant planets compared to previous studies. In the cases where multiple colors were reported across several nights for a single object, the colors were averaged. If colors were reported again in a later iteration of the survey, the most recent result was used. 
\label{cccomp}}
\end{figure}

The measured colors, as opposed to the absolute brightnesses, should be independent of assumptions made about the phase functions (provided  the latter are achromatic). The six panels in Figure (\ref{brbr}) compare  B-R values from this work and from previous surveys, with two panels included for the Jovian and Uranian satellites to compare colors from different authors.  Considered as a whole, the panels show that the colors are scattered on both sides of the diagonal line, as expected from random errors of measurement and/or rotational lightcurve variations.  There is a hint of a systematic error of unknown origin between the Rettig et al.~(2001) measurements of Jovian satellites and those from the present work. However, amounting to 0.05 to 0.10 magnitudes in B-R, we consider this unimportant (and perhaps not even statistically significant) compared to the much larger random errors indicated by the wide scatter of points about the diagonal line. This scatter largely reflects the difficulty of photometric measurements on faint satellites observed against the complex scattered light field from the nearby parent planet.  For example, the colors of bright object Lysithea (which has V $\sim$ 18) agree  with the data from both Rettig et al.~(2001) as well as Grav and Bauer (2007) to within $\sim \pm$ 0.01 magnitudes. The much fainter Stephano (V $\sim$ 25.4) shows color differences between this work and Grav et al.~(2004) of $\sim$ 0.70 magnitudes in B-V.  Figure (\ref{cccomp}) compares the color determined in this work with colors from published surveys.  
The final absolute magnitudes and colors are listed  in Table \ref{results}.

\section{Discussion}
\subsection{Colors}

\begin{figure}
\epsscale{0.85}
\plotone{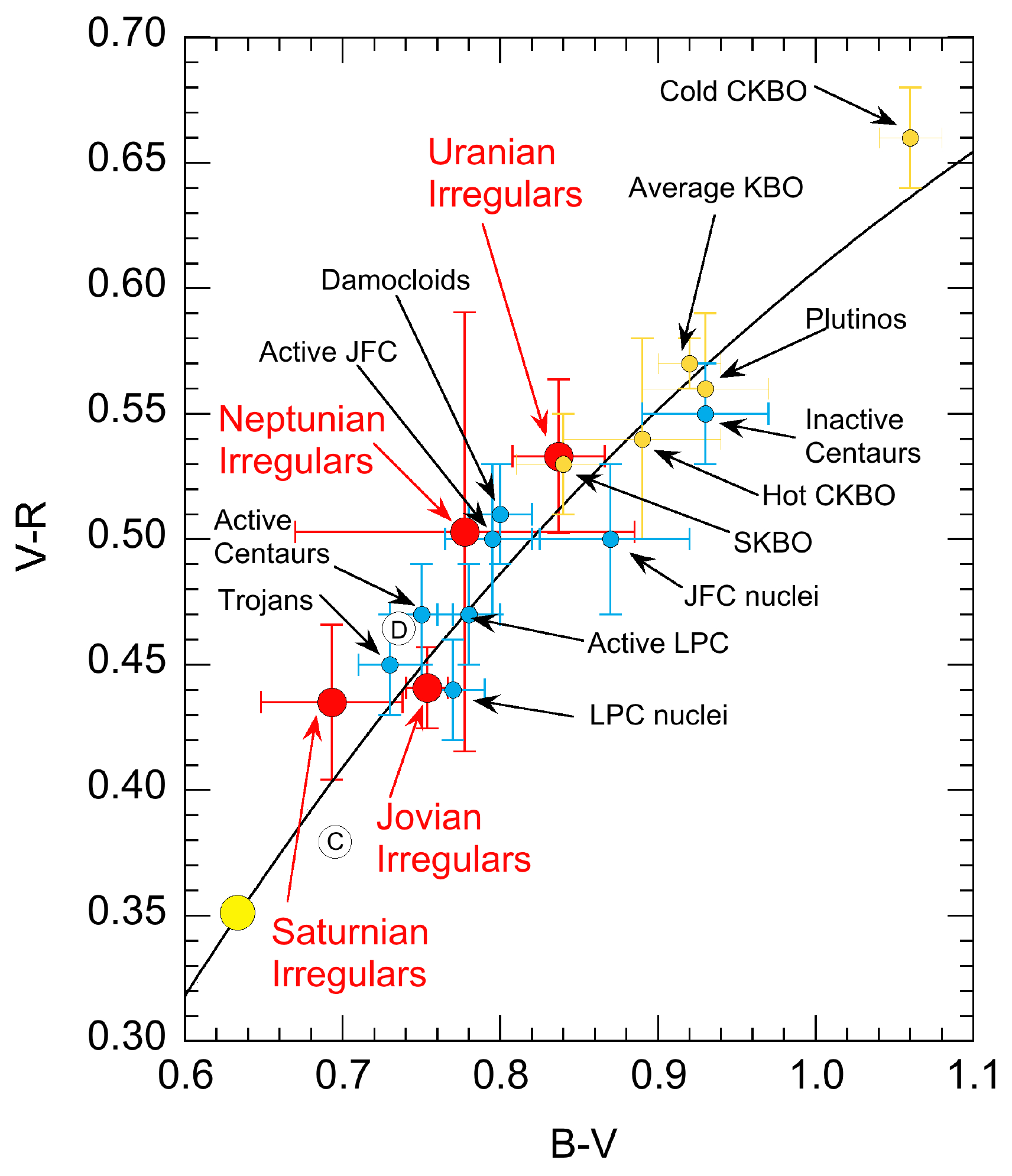}
\caption{ Color vs. color plot of small body populations in the solar system. Yellow data points represent Kuiper belt objects and blue data points represent comet, or comet-like objects. The circles labeled "C" and "D" represent the average color of the C-class and D-class asteroid populations respectively. The red data points are the average colors of the irregular satellites for each of the four giant planets. The color of the Sun is represented by the large yellow circle (Holmberg et al.~2006). 
\label{colorplot}}
\end{figure}

The irregular satellite colors were averaged at each planet and are plotted in Figure (\ref{colorplot}) together with the mean colors of other small body populations from Jewitt (2015).  The ``reddening line'' that spans Figure (\ref{colorplot}) (and is also present in the plots from Figure (\ref{cccomp})) from bottom left to upper right shows the locus of colors of objects having  linear normalized reflectivity gradients, $S'(\lambda)$ (measured in \%/1000 \AA), defined by $S'(\lambda) = (dS(\lambda)/d\lambda)/\overline{S} = $ constant, where $S(\lambda)$ is the ratio of the flux density at wavelength $\lambda$ to the flux density of the Sun and $\overline{S}$ is the average value of $S$ (Jewitt and Meech 1988). The reddening line is not a fit to the data and has no free parameters, other than being forced to pass through the B-V, V-R colors of the Sun. The figure shows that the satellite data all fall on the reddening line within the $\pm$1$\sigma$  error bars, indicating that they collectively possess linear reflectivity spectra as, indeed, do most objects in the outer solar system (Jewitt 2015).   The mean optical colors of the irregular satellite populations (B-V, V-R = 0.75$\pm$0.01, 0.44$\pm$0.02 at Jupiter, 0.69$\pm$0.04, 0.44$\pm$0.03 at Saturn, 0.84$\pm$0.03, 0.53$\pm$0.03 at Uranus, and  0.77$\pm$0.11, 0.50$\pm$0.09 at Neptune) are less red than either the hot (B-V, V-R = 0.89$\pm$0.05, 0.54$\pm$0.04) or cold (1.06$\pm$0.02, 0.66$\pm$0.02) components of the Kuiper belt, redder than the C-type asteroids ($\sim$0.70, $\sim$0.38; Dandy et al.~2003) but most similar to the D-type asteroids ($\sim$0.73, $\sim$0.46), as shown in Figure (\ref{colorplot}).  The D-types are especially abundant in the Jovian Trojan population but have a minor presence also in the main belt. 

\begin{figure}
\epsscale{0.75}
\plotone{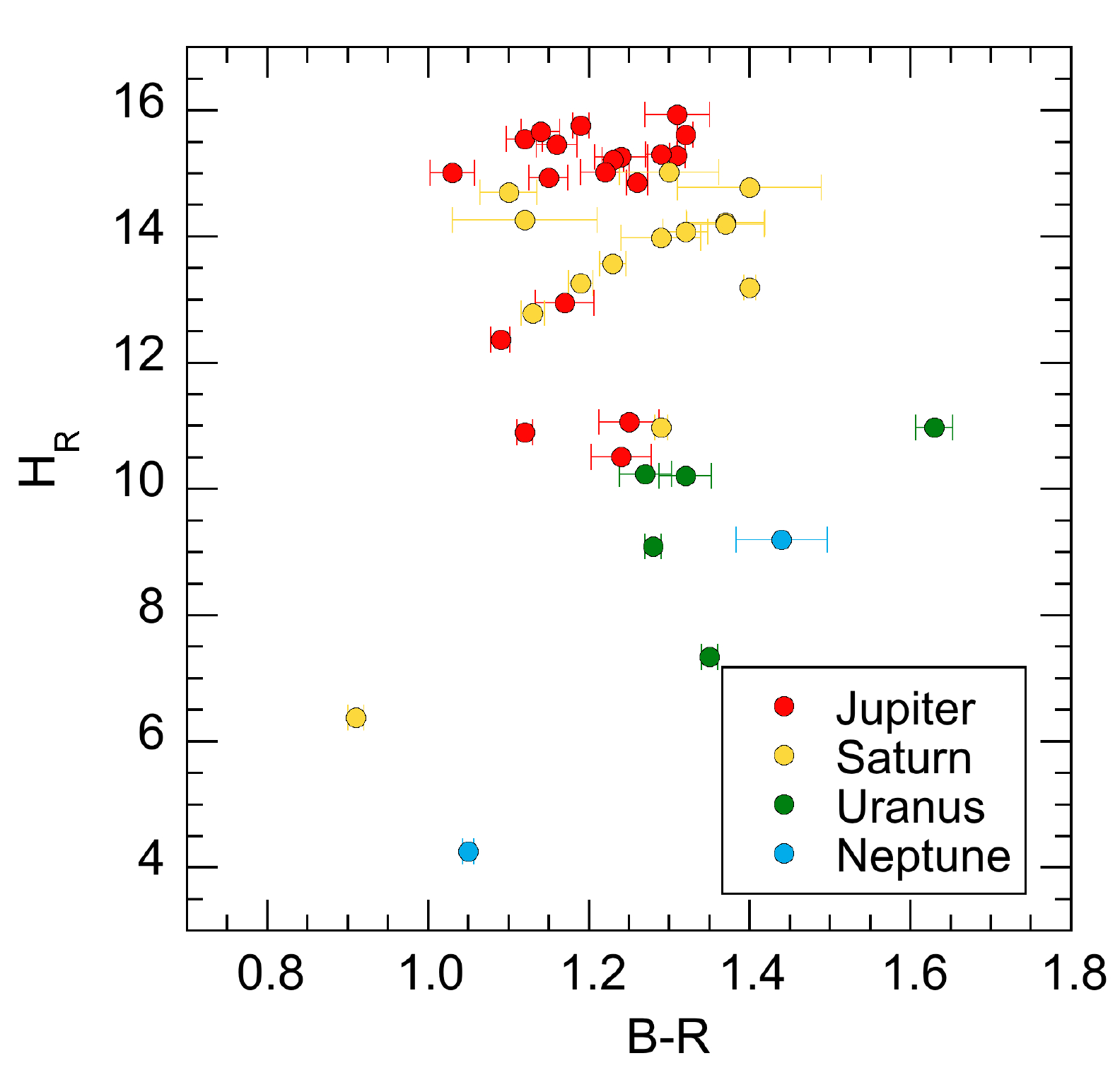}
\caption{B-R color versus absolute R magnitude. There are no apparent correlations, implying that the color does not depend on magnitude. 
\label{brvsR}}
\end{figure}

Figure (\ref{colorplot}) shows that the irregular satellites of Jupiter, Saturn, Neptune and Uranus are clustered near each other in color-color space, and are red-grey in color implying that the color of the irregular satellites does not depend on distance from the Sun. The color of each individual satellite is independent of its magnitude, as shown in Figure (\ref{brvsR}). This lack of dependency of magnitude and similarity in color further signifies a common origin. 

\begin{figure}
\epsscale{0.75}
\plotone{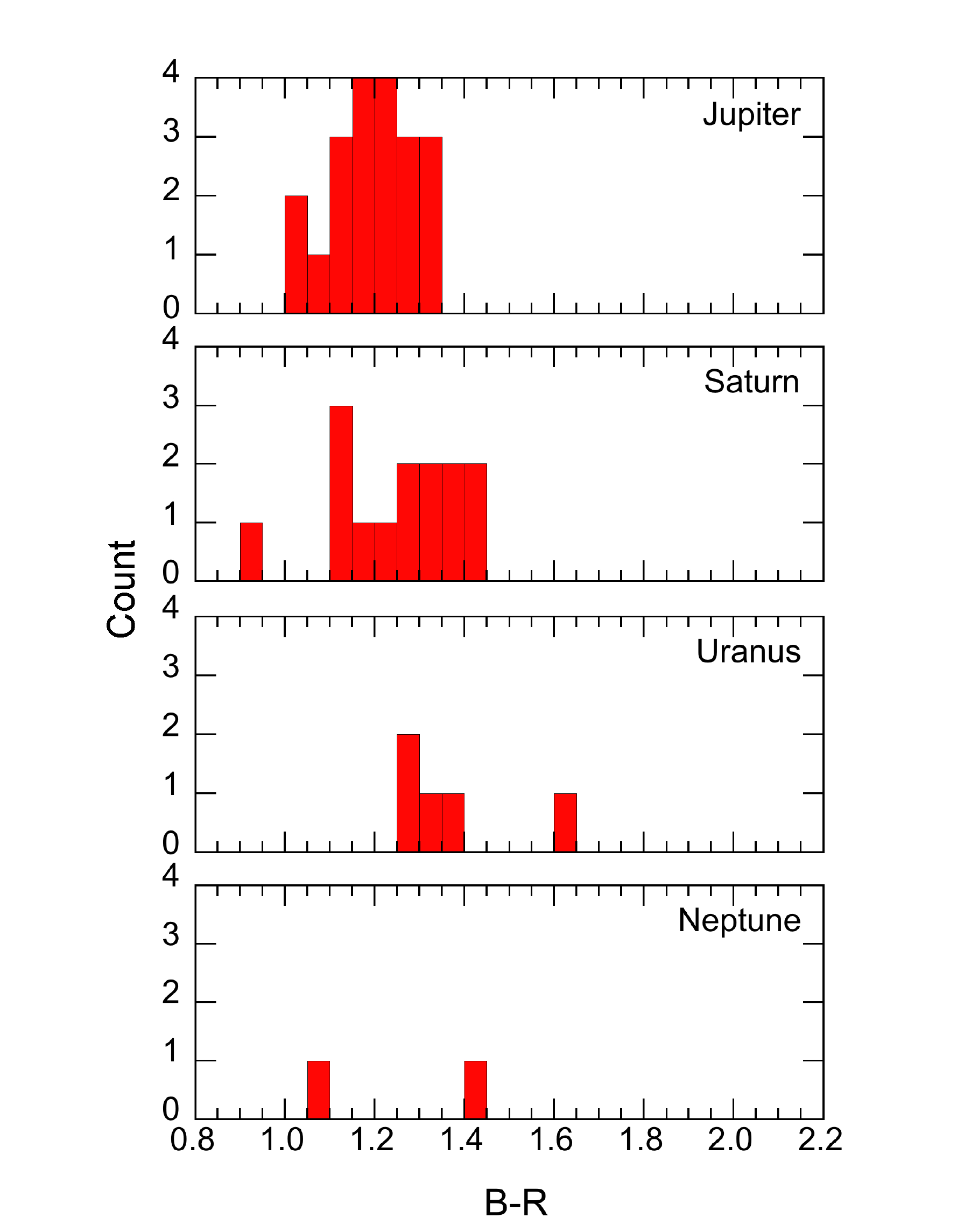}
\caption{  Histograms of B-R magnitudes of the irregular satellites observed in this survey at each of the giant planets.   Error bars on the colors are mostly comparable to, or smaller than, the bin size.
\label{histogram}}
\end{figure}

We used the Kolmogorov-Smirnov test as well as the Anderson-Darling (1954) test to assess the likelihood that the B-R colors of the different satellite populations could be drawn by chance from a common parent population. The Anderson-Darling test is more sensitive to the differences of the tails of the compared populations resulting mostly in  lower probabilities than produced by the Kolmogorov-Smirnov test. The most discrepant colors, as suggested visually from Figure (\ref{histogram}), occur between the Jupiter and Uranus satellite color distributions, which have probabilities of not sharing a common parent distribution of $\sim$99\%  according to the Kolmogorov-Smirnov test (Table \ref{KS}), and $\sim$99.5\% according to the Anderson-Darling test (Table \ref{AD}).  However, even the more stringent of these still does not meet the nominal 99.7\% probability associated with a 3$\sigma$ detection in a Gaussian distribution.  Given this, and the very small Uranus satellite sample size, we do not regard the difference as significant.  We also compared the irregular satellite colors with the Jovian Trojan color distribution (from Peixinho et al.~2015), finding no evidence for a significant difference. 


Jarvis et al. (2000) suggested that  asteroids ejected from the Kirkwood gaps in the main belt might have been captured by Jupiter.  Vilas et al.~(2006) reported spectral similarities between the irregular satellites and  main belt asteroids (specifically the C- and D-class asteroids in  the  classification system of  Tholen 1989) and, on this basis, also suggested that the main-belt is the  source region for the Jovian irregular satellites.  However, a main-belt source seems hard to support for two reasons. First, the numbers of Jovian Trojans and main-belt asteroids larger than 5 km in size are similar (Shoemaker et al. 1989 and Jewitt et al. 2000) but, while most Trojans are D- or P-types, such spectral classifications are rare in the main-belt asteroids.  Second, the asteroid belt at $\sim$2 to 3 AU hardly seems a good source for the irregular satellites of Saturn (10 AU), Uranus (20 AU) or Neptune (30 AU).

A currently popular suggested source region for the irregular satellites is the Kuiper belt, with the suggestion being that the irregular satellites could have been scattered from the Kuiper belt during planetary migration (Morbidelli et al.~2005). However, it is clear from Figure (\ref{colorplot}) that the average colors of the irregular satellites at each planet are all bluer than any of the Kuiper belt sub-populations, as are the Jovian Trojans.  If the irregular satellites were captured from a trans-Neptunian source, then their optical colors must have been modified after capture.  The specific difference is that the Kuiper belt population contains ultrared matter (B-R $>$ 1.6, Jewitt 2002) while only one irregular satellite (UXX Stephano, with B-R = 1.63$\pm$0.06, see Table \ref{results}) is marginally consistent with ultrared color.  Some evidence in seeming support of color modification is provided by observations of the Centaurs, which show a broad distribution of colors for large perihelion distances, $q \gtrsim$ 8 to 10 AU, but which lack ultrared members at smaller perihelion distances. Similarly, the  nuclei of Jupiter family comets  also lack ultrared matter, even though they were extracted from the Kuiper belt via the Centaurs. A plausible mechanism is resurfacing, caused by the ejection of particles at sub-orbital velocities in response to sublimation (Jewitt 2002). Also, the dynamical families of  irregular satellites are likely collisionally produced  (Nesvorny, 2003). Bottke et al.~(2013) explored the possibility that collisions between the irregular satellites cause dark material to be distributed onto the surfaces of the inner regular satellites. Though some of the dust from collisions can be lost, a portion  could fall back and cover the surface, similar to the slow, sublimated particles of the Centaurs and comets. Another possibility is a chemical change caused by volatilization of trace species as objects approach the Sun (Wong and Brown 2017). However, the critical distances for resurfacing (8 to 10 AU in the resurfacing hypothesis, where outgassing activity is first triggered by crystallization of amorphous ice; Jewitt 2002, 2009, 2015; and $\sim$17 AU in the H$_2$S model of Wong and Brown; 2017) are too small for the satellites of Uranus (at 20 AU) and Neptune (30 AU) to be affected. If Centaur-like color modification were the operative process, then we should expect to find ultrared matter in the satellites of Uranus and Neptune and, possibly even Saturn with an abundance $\sim$1/3, as in the dynamically hot populations of the Kuiper belt. A similar ``color conundrum" was recently identified in the colors of Jovian and Neptunian Trojans, which are similar to each other but unlike any plausible source population in the Kuiper belt (Jewitt 2018).

\subsection{Shapes}
\begin{figure}
\epsscale{1}
\plotone{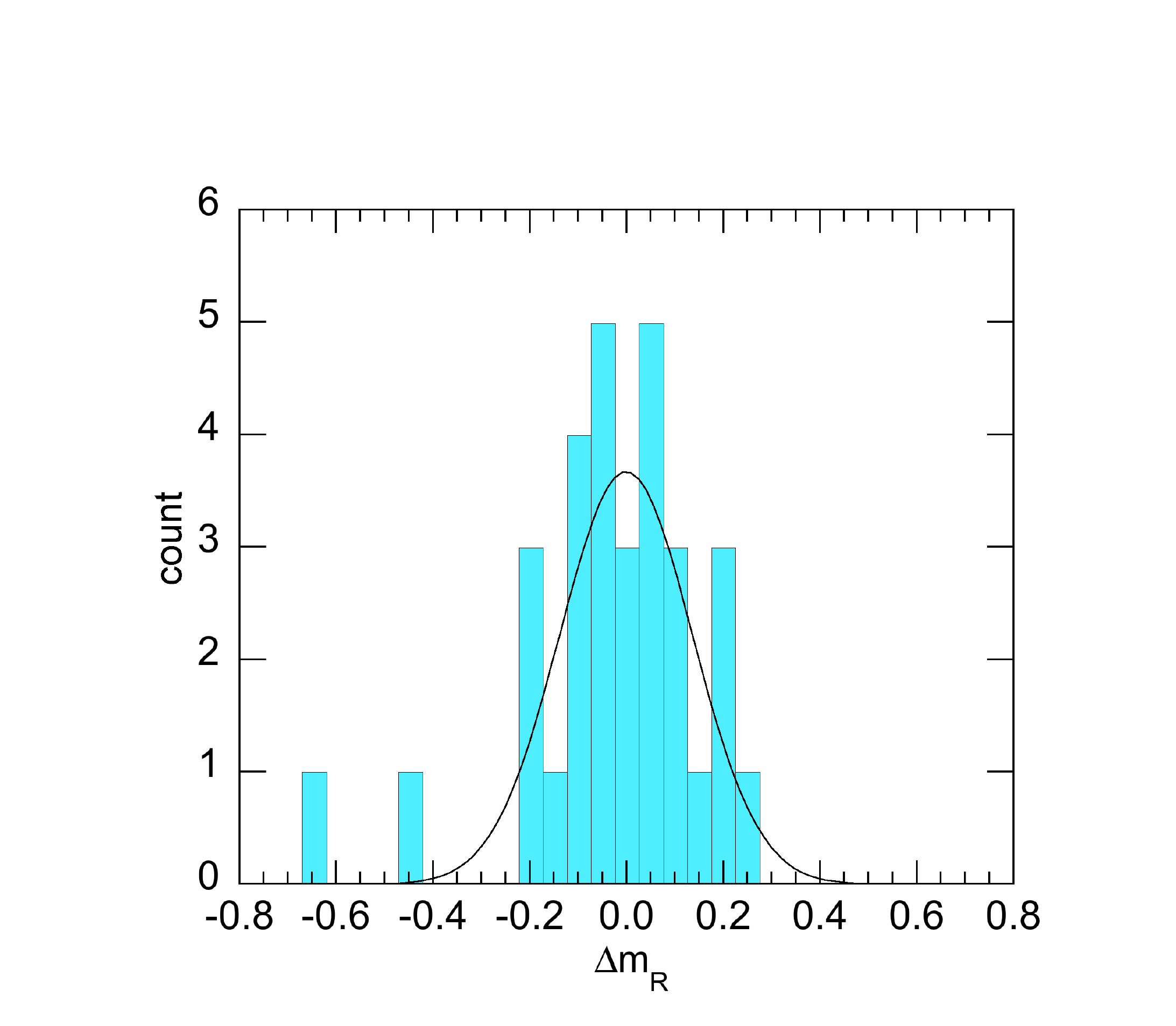}
\caption{ The distribution of differences in absolute magnitude of the irregular satellites measured on different days (blue histogram) compared  with a least-squares fitted Gaussian (black line). The width of the distribution indicates that the irregular satellites have a sky-plane axis ratio $b/a$ = 1.16,  similar to the mean projected shape of the main-belt asteroids. Collisional control is likely responsible for the shapes of objects of both types. 
\label{magdiff}}
\end{figure}

Our data also offer statistical information about the average shapes of the irregular satellites. Figure (\ref{magdiff}) shows the difference in the absolute magnitudes of satellites that were observed on two different days. The differences in magnitude do not depend on absolute magnitude. According to the Anderson-Darling test, the measured distribution of differences (blue histogram in the figure) is consistent with derivation from a Gaussian parent population (the probability that a larger Anderson-Darling statistic could be obtained by chance is 0.032).  The least-squares fit of a Gaussian is also shown in the Figure.   The fit has mean -0.001$\pm$0.003 magnitudes, consistent with zero, and FWHM = 0.32$\pm$0.01 magnitudes. (Using maximum likelihood estimation as an alternative, we obtained a fit with mean 0.006$\pm$0.02 magnitudes and FWHM = 0.28$\pm$0.02 magnitudes, consistent with the parameters found with the least-square fit.)

\begin{figure}
\epsscale{1}
\plotone{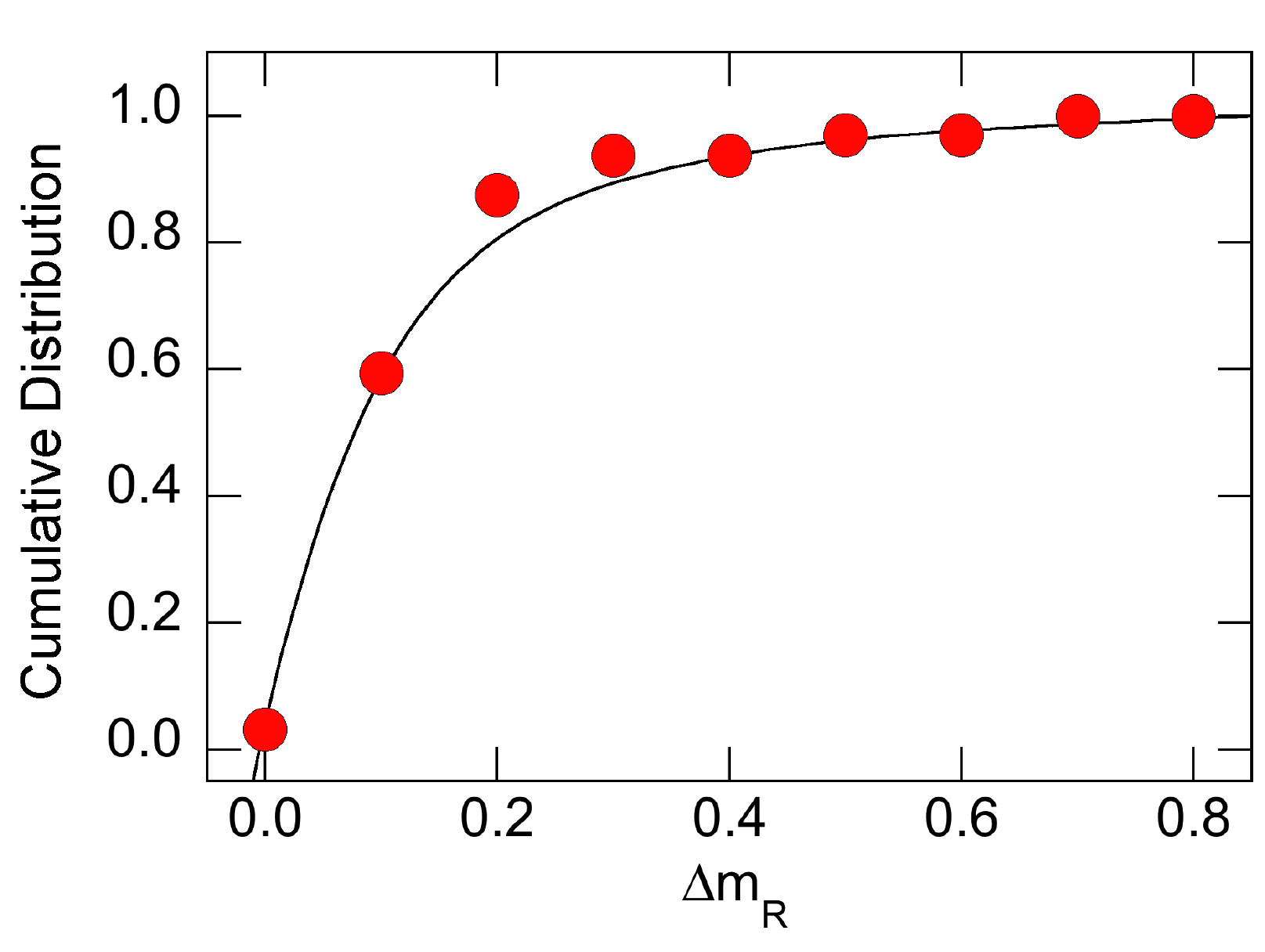}
\caption{Comparison of the normalized cumulative distribution of brightness differences of the irregular satellites from this work (red circles)  with  the same distribution for main-belt asteroids (black line) reported by Szab\'o and Kiss (2008). The  distributions match well, consistent with a common origin by collisions. 
\label{cumdist}}
\end{figure}

The shape can be estimated from the lightcurve range, $\Delta m_R$, using

\begin{equation}
\Delta m_R = 2.5\log({b/a})
\label{deltamr}
\end{equation}

\noindent where the body is taken to be elongated in shape with  long and short axes $b$ and $a$, respectively, both projected into the sky-plane.   We assume that the pair-wise observations of each satellite are uncorrelated with the rotational phase.  Then, our estimate of the average photometric range is $\Delta m_R$ = FWHM/2 = 0.16$\pm$0.01 magnitudes and substitution into Equation (\ref{deltamr}) gives  a sky-plane axis ratio $b/a$ = 1.16$\pm$0.01.   Szab\'o and Kiss (2008) made a statistical analysis of 11,735 asteroids and found that $b/a$ peaks at 1.2, with 80\% of the data falling in the range of $b/a$ = 1.1-1.2, which we regard as consistent with the average irregular satellite value. The normalized cumulative distributions of the brightness differences of the irregular satellites (red circles) are compared with those of  asteroids (black line) from Szab\'o and Kiss (2008) in Figure (\ref{cumdist}). We conclude that there is no observational evidence for a difference between the average shapes of the irregular satellites and the asteroids.  Given that the shapes of the asteroids are collisionally determined,  we likewise conclude that irregular satellites are also shaped by collisions, and this is consistent both with the existence of dynamical families in the Jovian satellite population, and with the inference by Bottke et al.~(2013) that irregular satellites are, as a group, highly collisionally processed.

\section{Conclusion}
 
We present the absolute magnitudes and colors of the irregular planetary satellites at each of the giant planets and use their average population colors to compare them to other populations in the solar system in search for a common origin. 
 
 \begin{itemize}
 \item The optical colors of the irregular satellites of the four giant planets are statistically similar to each other and independent of heliocentric distance. 
 
 \item The satellites lack the ultrared matter that colors the surfaces of many Kuiper belt objects.  About 80\% of the cold-classical and 30\% to 40\% of the hot classical Kuiper belt objects have B-R $>$ 1.60, whereas at most one of the measured irregular satellites (UXX Stephano, with B-R = 1.63$\pm$0.06) falls in the same range.
 
 
  \item If the irregular satellites were captured from the Kuiper belt, then their surface colors must have been modified.  The lack of ultrared surfaces even on the (cold) irregular satellites of Uranus and Neptune suggests that such modification cannot have been by any plausible thermal process. 
  
\item The means and the distributions of the shapes of the irregular satellites (average projected axis ratio  $b/a$ = 1.16$\pm$0.01) and main-belt asteroids ($b/a$ = 1.1-1.2, Szabo and Kiss 2008) are similar.  Collisional shattering likely determines the shapes of both types of object.  
 \end{itemize}

\acknowledgments

We thank Danielle Hastings, Dave Milewski, Man-To Hui and the anonymous referee for their comments. The data presented herein were obtained at the W. M. Keck Observatory, which is operated as a scientific partnership among the California Institute of Technology, the University of California and the National Aeronautics and Space Administration. The Observatory was made possible by the generous financial support of the W. M. Keck Foundation.   This work was supported, in part,  by a grant to DJ from NASA.

\clearpage

\begin{deluxetable}{lcc}
\tablecaption{Irregular Satellites
\label{number}}
\tablewidth{0pt}
\tablehead{ \colhead{\textbf{Planet}} & \colhead{\textbf{Known}} &  \colhead{\textbf{This Work}}  }
\startdata
 Jupiter & 60  & 20 \\
Saturn  & 38  & 14 \\
Uranus  & 9   & 6  \\
Neptune & 7   & 3  \\
\textbf{Total}   & 114 & 43
\enddata


\end{deluxetable}

 \clearpage

 \begin{landscape}

\begin{deluxetable}{lccccccccccccccccccccccccc}
\tabletypesize{\scriptsize}
\tablecaption{Geometry and Photometry
\label{geometry}}
\tablewidth{0pt}
\tablehead{ \colhead{Satellite} &  \colhead{UT Date and Time\tablenotemark{a}}   & $r_H$ (AU)\tablenotemark{b} & $\Delta$ (AU) \tablenotemark{c} & \colhead{$\alpha$ (\degr) \tablenotemark{d}} &  \colhead{ R\tablenotemark{e}}  & \colhead{B-V} & \colhead{V-R}   & \colhead{B-R} }
\startdata
\textbf{Jupiter}                \\

JIX Sinope         & 2008 Sep 30  05:43-05:48 & 5.00    & 4.80       & 11.50                 & 18.42      $\pm$ 0.07 & 0.77 $\pm$ 0.08 & 0.48 $\pm$ 0.10 & 1.25 $\pm$ 0.08 \\
JX Lysithea        & 2008 Sep 30 05:20-05:30  & 5.04    & 4.04       & 1.36                  & 17.50      $\pm$ 0.01     & 0.72 $\pm$ 0.01     & 0.40 $\pm$ 0.01     & 1.12 $\pm$ 0.01     \\
JXI Carme          & 2008 Sep 30 05:35-05:43  & 5.04  & 4.85  & 11.41 & 17.92 $\pm$  0.07 & 0.76  $\pm$  0.08 & 0.48  $\pm$  0.08 & 1.24  $\pm$  0.05 \\ 
JXIII Leda         & 2009 Aug 19 07:28-07:43  & 5.05    & 4.04       & 1.32                  & 19.03      $\pm$ 0.01     & 0.69 $\pm$ 0.01     & 0.36 $\pm$ 0.01     & 1.05 $\pm$ 0.01     \\
                   & 2009 Aug 21 13:37-13:44  & 5.05    & 4.05       & 1.89                  & 18.84      $\pm$ 0.01     &  $\dots$     &    $\dots$          & 1.20 $\pm$ 0.02     \\
JXVIII Themisto    & 2008 Sep 30 05:50-06:05  & 5.07    & 4.06       & 1.03                  & 19.48      $\pm$ 0.01     & 0.79 $\pm$ 0.02     & 0.52 $\pm$ 0.02     & 1.32 $\pm$ 0.01     \\
                   & 2008 Sep 30 06:25-06:36  & 5.06    & 4.06       & 1.54                  & 19.78      $\pm$ 0.04     &   $\dots$    &  $\dots$            & 0.83 $\pm$ 0.04     \\
                   & 2009 Aug 19 07:43-07:49  & 5.06    & 4.06       & 1.57                  & 19.45      $\pm$ 0.05     &   $\dots$    &    $\dots$         & 1.22 $\pm$ 0.06     \\
                   & 2009 Aug 21 07:14-70:28  & 5.06    & 4.06       & 1.60                  & 19.58      $\pm$ 0.01     &    $\dots$   &  $\dots$            & 1.31 $\pm$ 0.01     \\
JXIX Megaclite     & 2009 Aug 19 09:49-09:55  & 5.15    & 4.14       & 1.25                  & 21.55      $\pm$ 0.01     & 0.81 $\pm$ 0.07     & 0.44 $\pm$ 0.07     & 1.26 $\pm$ 0.02     \\
                   & 2009 Aug 21 07:28-07:32  & 5.15    & 4.15       & 1.77                  & 21.52      $\pm$ 0.04     &  $\dots$     &            & 1.23 $\pm$ 0.07     \\
                   & 2009 Aug 21 10:33-10:43  & 5.15    & 4.15       & 1.79                  & 21.76      $\pm$ 0.01     & $\dots$      &               & 0.92 $\pm$ 0.03     \\
                   & 2009 Aug 21 11:09-11:19  & 5.15    & 4.15       & 1.81                  & 21.76      $\pm$ 0.02     & $\dots$      &          $\dots$      & 1.38 $\pm$ 0.05     \\
JXX Taygete        & 2009 Aug 19 07:57-08:06  & 5.08    & 4.07       & 0.86                  & 21.88      $\pm$ 0.02     & 0.77 $\pm$ 0.03     & 0.48 $\pm$ 0.02     & 1.26 $\pm$ 0.03     \\
                   & 2009 Aug 21 13:26-13:37  & 5.08    & 4.08       & 1.41                  & 21.95      $\pm$ 0.01     &    $\dots$   &   $\dots$       & 1.35 $\pm$ 0.03     \\
JXXI Chaldene      & 2009 Aug 19 08:16-08:27  & 5.06    & 4.05       & 0.57                  & 22.19      $\pm$ 0.01     & 0.82 $\pm$ 0.05     & 0.50 $\pm$ 0.05     & 1.32 $\pm$ 0.03     \\
JXXII Harpalyke    & 2009 Aug 19 08:59-09:08  & 4.92    & 3.91       & 1.30                  & 22.02      $\pm$ 0.02     & 0.70 $\pm$ 0.02     & 0.42 $\pm$ 0.03     & 1.12 $\pm$ 0.02     \\
JXXIII Kalyke      & 2009 Aug 19 09:17-09:27  & 5.15    & 4.14       & 0.67                  & 21.61      $\pm$ 0.02     & 0.69 $\pm$ 0.03     & 0.46 $\pm$ 0.03     & 1.15 $\pm$ 0.03     \\
JXXIV Iocaste      & 2009 Aug 19 09:36-09:42  & 4.93    & 3.92       & 0.95                  & 21.72      $\pm$ 0.03     & 0.86 $\pm$ 0.08     & 0.38 $\pm$ 0.07     & 1.24 $\pm$ 0.05     \\
JXXV Erinome      
                   & 2009 Aug 21 12:06-12:13  & 4.89    & 3.89       & 1.45                  & 22.11      $\pm$ 0.04     & 0.72 $\pm$ 0.01     & 0.43 $\pm$ 0.04     & 1.14 $\pm$ 0.04     \\
JXXVI Isonoe       & 2009 Aug 19 10:59-11:16  & 5.14    & 4.14       & 1.46                  & 22.63      $\pm$ 0.04     & 0.78 $\pm$ 0.05     & 0.53 $\pm$ 0.04     & 1.31 $\pm$ 0.06     \\
JXXVII Praxidike   & 2009 Aug 19 11:16-11:36  & 4.93    & 3.92       & 1.10                  & 21.48      $\pm$ 0.03     & 0.71 $\pm$ 0.05     & 0.32 $\pm$ 0.05     & 1.03 $\pm$ 0.03     \\
JXXVIII Autonoe    & 2009 Aug 19 11:36-11:58  & 4.97    & 3.96       & 1.32                  & 21.74      $\pm$ 0.02     & 0.81 $\pm$ 0.02     & 0.48 $\pm$ 0.03     & 1.29 $\pm$ 0.03     \\
                   & 2009 Aug 21 10:12-10:22  & 4.96    & 3.96       & 1.83                  & 21.70      $\pm$ 0.03     &    $\dots$   &   $\dots$            & 1.25 $\pm$ 0.04     \\
                   & 2009 Aug 21 11:26-11:39  & 4.96    & 3.96       & 1.84                  & 21.76      $\pm$ 0.03     &    $\dots$   &     $\dots$             & 1.15 $\pm$ 0.06     \\
                   & 2009 Aug 21 12:20-12:42  & 4.96    & 3.96       & 1.86                  & 21.79      $\pm$ 0.02     & 0.73 $\pm$ 0.14 & 0.50 $\pm$ 0.07 & 1.22 $\pm$ 0.12     \\
JXXIX Thyone       & 2009 Aug 19 11:58-12:28  & 5.12    & 4.11       & 0.90                  & 22.10      $\pm$ 0.03     & 0.71 $\pm$ 0.06     & 0.46 $\pm$ 0.04     & 1.16 $\pm$ 0.05     \\
JXXX Hermippe      & 2009 Aug 19 12:28-12:49  & 5.03    & 4.03       & 0.93                  & 21.59      $\pm$ 0.03     & 0.72 $\pm$ 0.05     & 0.49 $\pm$ 0.04     & 1.22 $\pm$ 0.06     \\
JXVII Callirrhoe   & 2009 Aug 19 13:21-13:30  & 5.16    & 4.15       & 1.20                  & 20.90      $\pm$ 0.07     & 0.80 $\pm$ 0.11     & 0.24 $\pm$ 0.12     & 1.04 $\pm$ 0.08     \\
JXLVII Eukelade    & 2009 Aug 21 12:52-13:20  & 4.94    & 3.94       & 1.25                  & 21.79      $\pm$ 0.02     & 0.78 $\pm$ 0.07 & 0.50 $\pm$ 0.07 & 1.29 $\pm$ 0.02     \\
JXLVIII Cyllene    & 2009 Aug 21 11:39-12:06  & 5.03    & 4.02       & 1.51                  & 22.35      $\pm$ 0.01     & 0.73 $\pm$ 0.07     & 0.47 $\pm$ 0.07     & 1.19 $\pm$ 0.01     \\
 \\
\midrule
\\
\textbf{Saturn}   &                          &       &           \\
SIX Phoebe         & 2008 Mar 10 09:19-09-27  & 9.28    & 8.33       & 1.89                  & 15.88      $\pm$ 0.01     & 0.58 $\pm$ 0.01 & 0.34 $\pm$ 0.01 & 0.91 $\pm$ 0.01     \\
SXXI Tarvos        & 2008 Mar 10 10:33-11:13  & 9.25    & 8.30       & 1.84                  & 22.28      $\pm$ 0.01     & 0.71 $\pm$ 0.04     & 0.42 $\pm$ 0.03     & 1.13 $\pm$ 0.03     \\
SXXII Ijiraq       & 2008 Mar 11 06:49-07:19  & 9.31    & 8.36       & 1.94                  & 22.73      $\pm$ 0.01     &    $\dots$   &      $\dots$          & 1.40 $\pm$ 0.03     \\
SXXVI Albiorix     
                   & 2008 Mar 10 08:26-09:19  & 9.25    & 8.25       & 0.34                  & 20.43      $\pm$ 0.01     & 0.75 $\pm$ 0.02     & 0.48 $\pm$ 0.01     & 1.23 $\pm$ 0.02     \\
                   & 2008 Mar 10 11:13-11:25  & 9.28    & 8.33       & 1.88                  & 20.48      $\pm$ 0.03     &       $\dots$      &    $\dots$       & 1.23 $\pm$ 0.06     \\
                   & 2008 Mar 11 06:07-06:28  & 9.28    & 8.33       & 1.96                  & 20.46      $\pm$ 0.01     & 0.91 $\pm$ 0.02     & 0.50 $\pm$ 0.02     & 1.41 $\pm$ 0.01     \\
SXXVIII Erriapus   & 2008 Mar 10 08:11-08:26  & 9.33    & 8.34       & 0.33                  & 22.72      $\pm$ 0.02     & 0.77 $\pm$ 0.01  & 0.42 $\pm$ 0.02 & 1.19 $\pm$ 0.02     \\
SXXXI Narvi        & 2008 Mar 11 06:28-06:33  & 9.34    & 8.38       & 1.81                  & 23.52      $\pm$ 0.05     &  $\dots$     &     $\dots$          & 1.29 $\pm$ 0.08     \\
SXXXVII Bebhionn   & 2008 Mar 10 09:40-10:05  & 9.30    & 8.34       & 1.77                  & 23.78      $\pm$ 0.18     & 0.60 $\pm$ 0.10     & 0.51 $\pm$ 0.20     & 1.12 $\pm$ 0.18     \\
                   & 2008 Mar 11 08:40-08:45  & 9.29    & 8.34       & 1.87                  & 23.74      $\pm$ 0.02     &  $\dots$     &    $\dots$          & 1.14 $\pm$ 0.05     \\
SXXXVI Aegir       & 2008 Mar 11 07:19-07:24  & 9.26    & 8.14       & 2.03                  & 24.49      $\pm$ 0.06     & $\dots$      &       $\dots$       & 1.30 $\pm$ 0.06     \\
SXXXVIII Bergelmir & 2008 Mar 11 12:33-12:38  & 9.40    & 8.45       & 1.93                  & 24.28      $\pm$ 0.04     &   $\dots$    & $\dots$           & 1.10 $\pm$ 0.15     \\
SXXXIX Bestla      & 2008 Mar 11 10:15-10:20  & 9.18    & 8.24       & 2.06                  & 23.55      $\pm$ 0.03     &     $\dots$  &    $\dots$          & 1.32 $\pm$ 0.04     \\
SXLII Fornjot      & 2008 Mar 11 08:05-08:10  & 9.37    & 8.42       & 2.02                  & 24.34      $\pm$ 0.09     &  $\dots$     &        $\dots$       & 1.40 $\pm$ 0.09     \\
SLII Tarqeq        & 2008 Mar 11 11:45-11:50  & 9.34    & 8.39       & 1.94                  & 23.12      $\pm$ 0.02     &   $\dots$    &       $\dots$         & 1.23 $\pm$ 0.07     \\
S/2007 S2         & 2008 Mar 11 11:55-12:30     & 9.27    & 8.33       & 2.01                  & 23.74      $\pm$ 0.05     &  $\dots$     &  $\dots$              & 1.37 $\pm$ 0.06     \\
S/2007 S2         & 2008 Mar 11 11:55-12:30     & 9.30    & 8.40       & 2.00                  & 23.74      $\pm$ 0.05     & $\dots$      &         $\dots$        & 1.37 $\pm$ 0.09     \\
 \\
\midrule
\\
\textbf{Uranus}  &                          &       &       &       \\
UXVI Caliban       & 2008 Sep 05 10:25-11:29  & 20.15   & 19.15      & 0.32                  & 21.98      $\pm$ 0.01     & 0.87 $\pm$ 0.03     & 0.51 $\pm$ 0.01     & 1.39 $\pm$ 0.03     \\
                   & 2015 Dec 08 10:01-10:17  & 19.95   & 19.44      & 2.45                  & 22.17      $\pm$ 0.01     & 0.73 $\pm$ 0.01     & 0.43 $\pm$ 0.01     & 1.16 $\pm$ 0.01     \\
UXVII Sycorax      & 2008 Sep 04 11:29-11:47  & 20.06   & 19.05      & 0.08                  & 20.17      $\pm$ 0.02     & 0.77 $\pm$ 0.02     & 0.52 $\pm$ 0.02     & 1.29 $\pm$ 0.03     \\
                   & 2008 Sep 05 09:21-10:08  & 20.06   & 19.05      & 0.28                  & 20.24      $\pm$ 0.01     & 0.80 $\pm$ 0.01     & 0.56 $\pm$ 0.01     & 1.37 $\pm$ 0.01     \\
                   & 2015 Dec 08 08:16-08:33  & 20.01   & 19.48      & 2.44                  & 20.50      $\pm$ 0.01     & 0.85 $\pm$ 0.02     & 0.53 $\pm$ 0.02     & 1.38 $\pm$ 0.01     \\
UXVIII Prospero    & 2008 Sep 04 11:55-12:11  & 20.21   & 19.21      & 0.32                  & 23.18      $\pm$ 0.09     & 0.62 $\pm$ 0.03     & 0.53 $\pm$ 0.09     & 1.15 $\pm$ 0.09     \\
                   & 2008 Sep 30 11:13-11:59  & 20.21   & 19.26      & 0.90                  & 23.26      $\pm$ 0.04     & 0.96 $\pm$ 0.08     & 0.31 $\pm$ 0.08     & 1.26 $\pm$ 0.05     \\
                   & 2015 Dec 09 8:08-8:50    & 20.02   & 19.53      & 2.48                  & 23.20      $\pm$ 0.01     & 0.86 $\pm$ 0.06     & 0.67 $\pm$ 0.06     & 1.54 $\pm$ 0.02     \\
UXIX Setebos       & 2008 Sep 05 11:29-12:06  & 20.16   & 19.15      & 0.31                  & 23.17      $\pm$ 0.04     & 0.78 $\pm$ 0.02     & 0.49 $\pm$ 0.04     & 1.27 $\pm$ 0.04     \\

UXX Stephano       & 2008 Sep 04 12:31-13:24  & 20.15   & 19.15      & 0.26                  & 24.03       $\pm$  0.02     & 0.91 $\pm$ 0.12     & 0.71 $\pm$ 0.05     & 1.62 $\pm$ 0.11     \\
                   & 2008 Sep 05 12:12-12:39  & 20.15   & 19.15      & 0.35                  & 23.80      $\pm$  0.05     & 1.03 $\pm$ 0.09     & 0.61 $\pm$ 0.09     & 1.63 $\pm$ 0.07     \\
UXXI Trinculo      & 2008 Sep 05 13:53-14:22  & 20.04   & 19.04      & 0.34                  & 25.20      $\pm$  0.15     &  $\dots$  &  $\dots$  &  $\dots$   \\   
 \\
\midrule
 \\  \\  \\ \\
\textbf{Neptune} &                          &       &       &       \\
NI Halimede        & 2008 Sep 04 09:39-10:22  & 30.06   & 29.13      & 0.73                  & 23.72      $\pm$ 0.08     & 0.84 $\pm$ 0.18     & 0.72 $\pm$ 0.19     & 1.57 $\pm$ 0.09     \\
                   & 2008 Sep 05 07:47-08:49  & 30.06   & 29.13      & 0.73                  & 24.16      $\pm$ 0.08     & 0.92 $\pm$ 0.10     & 0.38 $\pm$ 0.12     & 1.31 $\pm$ 0.09     \\
NII Nereid         & 2008 Sep 04 06:36-06:57  & 30.02   & 29.07      & 0.66                  & 18.94      $\pm$ 0.01     & 0.65 $\pm$ 0.02     & 0.39 $\pm$ 0.01     & 1.04 $\pm$ 0.02     \\
                   & 2008 Sep 05 06:21-06:40  & 30.02   & 29.08      & 0.69                  & 19.04      $\pm$ 0.01     & 0.69 $\pm$ 0.01     & 0.37 $\pm$ 0.02     & 1.06 $\pm$ 0.01     \\
NIV Neso           & 2008 Sep 05 06:45-07:47  & 30.01   & 29.07      & 0.66                  & 24.66      $\pm$ 0.04     &       $\dots$        & 0.84 $\pm$ 0.12     &       $\dots$         \\
                   & 2008 Sep 30 08:54-10:12  & 30.01   & 29.31      & 1.38                  & 25.34      $\pm$ 0.31     &      $\dots$         & 0.31 $\pm$ 0.42     &   $\dots$

\enddata


\tablenotetext{a}{UT date and range of start times of the integrations}
\tablenotetext{b}{Heliocentric Distance in AU}
\tablenotetext{c}{Geocentric Distance in AU}
\tablenotetext{d}{Phase Angle in degrees}
\tablenotetext{e}{Apparent magnitude in the R filter}

\end{deluxetable}

\end{landscape}

\clearpage

\begin{deluxetable}{lcccc}
\tabletypesize{\scriptsize}
\tablecaption{Adopted Absolute Magnitudes and Colors
\label{results}}
\tablewidth{0pt}
\tablehead{  \colhead{Satellite} &  \colhead{$H_R$}  & \colhead{B-V} & \colhead{V-R} &\colhead{B-R}  }
\startdata
\textbf{Jupiter} &       &       &      &       \\
JIX Sinope         & 11.06  $\pm$  0.04 & 0.77  $\pm$  0.07 & 0.48  $\pm$  0.08 & 1.25  $\pm$  0.05 \\
JX Lysithea        & 10.97  $\pm$  0.01 & 0.72  $\pm$ 0.01 & 0.41  $\pm$  0.01 & 1.13 $\pm$  0.01 \\
JXI Carme          & 10.51  $\pm$  0.04 & 0.76  $\pm$  0.08 & 0.48  $\pm$  0.08 & 1.24  $\pm$  0.05 \\
JXIII Leda         & 12.36  $\pm$  0.01 & 0.66  $\pm$  0.01 & 0.43  $\pm$  0.01 & 1.09  $\pm$  0.01 \\
JXVIII Themisto    & 12.86  $\pm$  0.04 & 0.80  $\pm$  0.07 & 0.48 $\pm$ 0.08 & 1.28  $\pm$ 0.04 \\
JXIX Megaclite     & 14.85  $\pm$  0.01 & 0.82  $\pm$  0.07 & 0.44  $\pm$  0.07 & 1.26  $\pm$ 0.03 \\
JXX Taygete        & 15.28  $\pm$  0.01 & 0.84  $\pm$  0.02 & 0.47 $\pm$  0.02 & 1.31  $\pm$ 0.02 \\
JXXI Chaldene      & 15.61  $\pm$  0.01 & 0.82  $\pm$  0.05 & 0.50 $\pm$  0.05 & 1.32 $\pm$  0.03 \\
JXXII Harpalyke    & 15.54  $\pm$  0.02 & 0.70  $\pm$ 0.02 & 0.42 $\pm$  0.03 & 1.12  $\pm$  0.02 \\
JXXIII Kalyke      & 14.93  $\pm$  0.02 & 0.69  $\pm$  0.03 & 0.46 $\pm$  0.03 & 1.15  $\pm$  0.03 \\
JXXIV Iocaste      & 15.26 $\pm$  0.03 & 0.86 $\pm$  0.08 & 0.38  $\pm$  0.07 & 1.24  $\pm$  0.05 \\
JXXV Erinome       & 15.66  $\pm$  0.02 & 0.72  $\pm$  0.06 & 0.42  $\pm$  0.04 & 1.14  $\pm$  0.06 \\
JXXVI Isonoe       & 15.93  $\pm$  0.04 & 0.78  $\pm$  0.05 & 0.53  $\pm$  0.04 & 1.31  $\pm$  0.06 \\
JXXVII Praxidike   & 15.01  $\pm$  0.03 & 0.71  $\pm$  0.05 & 0.32  $\pm$  0.05 & 1.03  $\pm$  0.03 \\
JXXVIII Autonoe    & 15.21 $\pm$  0.01 & 0.72  $\pm$  0.03 & 0.51  $\pm$  0.02 & 1.23  $\pm$  0.04 \\
JXXIX Thyone       & 15.46  $\pm$  0.03 & 0.71  $\pm$  0.06 & 0.45  $\pm$  0.04 & 1.16  $\pm$  0.05 \\
JXXX Hermippe      & 15.02 $\pm$  0.03 & 0.73  $\pm$  0.05 & 0.49  $\pm$  0.04 & 1.22  $\pm$  0.06 \\
JXVII Callirrhoe   & 14.20  $\pm$  0.07 & 0.81  $\pm$  0.11 & 0.23  $\pm$  0.12 & 1.04  $\pm$  0.08 \\
JXLVII Eukelade    & 15.30  $\pm$ 0.02 & 0.79 $\pm$  0.07 & 0.50 $\pm$  0.07 & 1.29  $\pm$  0.02 \\
JXLVIII Cyllene    & 15.76  $\pm$  0.01 & 0.73  $\pm$  0.07 & 0.46  $\pm$  0.07 & 1.19  $\pm$  0.01 \\
\midrule
\textbf{Saturn}  &       &       &      &       \\
SIX Phoebe         & 6.37   $\pm$  0.01 & 0.57  $\pm$  0.01 & 0.34  $\pm$  0.01 & 0.91  $\pm$  0.01 \\
SXXI Tarvos        & 12.78  $\pm$  0.01 & 0.71  $\pm$  0.04 & 0.42  $\pm$  0.03 & 1.13 $\pm$  0.03 \\
SXXII Ijiraq       & 13.19  $\pm$  0.01 &      $\dots$  &      $\dots$      & 1.40  $\pm$  0.02 \\
SXXVI Albiorix     & 10.97  $\pm$ 0.01 & 0.80  $\pm$  0.02 & 0.50  $\pm$ 0.01 & 1.29  $\pm$  0.02 \\
SXXVIII Erriapus   & 13.26  $\pm$  0.02 & 0.78  $\pm$  0.10 & 0.41  $\pm$  0.10 & 1.19  $\pm$ 0.02 \\
SXXXI Narvi        & 13.98 $\pm$  0.05 &      $\dots$      &   $\dots$     & 1.29  $\pm$ 0.08 \\
SXXXVII Bebhionn   & 14.26  $\pm$  0.09 & 0.61  $\pm$  0.10 & 0.51  $\pm$  0.13 & 1.12  $\pm$  0.09 \\
SXXXVI Aegir       & 15.02  $\pm$  0.06 &       $\dots$    &       $\dots$ & 1.30  $\pm$ 0.06 \\
SXXXVIII Bergelmir & 14.70  $\pm$  0.04 &    $\dots$    &     $\dots$         & 1.10  $\pm$  0.15 \\
SXXXIX Bestla      & 14.07  $\pm$  0.03 &    $\dots$     &    $\dots$    & 1.32  $\pm$  0.04 \\
SXLII Fornjot      & 14.77  $\pm$  0.09 &     $\dots$   &     $\dots$     & 1.40  $\pm$  0.09 \\
SXLIV Hyrrokkin    & 13.57  $\pm$  0.01 &       $\dots$       &    $\dots$    & 1.23  $\pm$  0.07 \\
SLII Tarqeq        & 14.22  $\pm$  0.05 &     $\dots$   &     $\dots$       & 1.37  $\pm$  0.06 \\
S/2007 S2         & 14.19 $\pm$  0.05 &    $\dots$    &     $\dots$       & 1.37  $\pm$  0.09 \\
\midrule
\textbf{Uranus}  &       &       &      &       \\
UXVI Caliban       & 9.09  $\pm$  0.01 & 0.81  $\pm$  0.02 & 0.47  $\pm$  0.02 & 1.28  $\pm$  0.02 \\
UXVII Sycorax      & 7.34   $\pm$  0.01 & 0.81  $\pm$  0.01 & 0.54  $\pm$  0.01 & 1.35  $\pm$ 0.01 \\
UXVIII Prospero    & 10.21  $\pm$  0.03 & 0.81  $\pm$  0.04 & 0.51  $\pm$  0.05 & 1.32  $\pm$  0.04 \\
UXIX Setebos       & 10.18  $\pm$  0.03 & 0.78 $\pm$ 0.02     & 0.49 $\pm$ 0.04     & 1.27 $\pm$ 0.04 \\
UXX Stephano       & 10.97  $\pm$  0.02 & 0.97  $\pm$  0.07 & 0.66  $\pm$  0.05 & 1.63  $\pm$  0.06 \\
UXXI Trinculo      & 12.28  $\pm$  0.15 &  $\dots$  &  $\dots$  &  $\dots$    \\
\midrule
\textbf{Neptune} &       &       &      &         \\
NI Halimede        & 9.20   $\pm$  0.06 & 0.89  $\pm$  0.10 & 0.56  $\pm$  0.11 & 1.44  $\pm$  0.06 \\
NII Nereid         & 4.26   $\pm$ 0.01 & 0.67  $\pm$  0.02 & 0.38  $\pm$  0.01 & 1.05  $\pm$  0.01 \\
NIV Neso           & 10.25 $\pm$  0.10 &      $\dots$   & 0.58  $\pm$  0.13 &      $\dots$ 

\enddata


\end{deluxetable}
\clearpage


\begin{deluxetable}{llcrrrccccr}

\tablecaption{Kolmogorov-Smirnov Probabilities\tablenotemark{a}
\label{KS}}
\tablewidth{0pt}
\tablehead{ \colhead{Group} &  \colhead{Jsat}  & \colhead{Ssat}  & \colhead{Usat} & \colhead{Nsat} & \colhead{JTro}   }
\startdata
Jsat   	& 1.000  &   0.300  & 0.010 &    0.585  & 0.848  \\
Ssat   	&  & 1.000 & 0.395 & 0.605 &  0.100 \\
Usat  &  &  & 1.000 & 0.705  & 0.007 \\
Nsat   &    &   &   &  1.000  & 0.657  \\
Jtro        &    &   &   &    & 1.000\\

\enddata

\tablenotetext{a}{Probability that any two given color distributions could be drawn from the same parent population.   The lower half of the diagonally symmetric matrix is not shown.}

\end{deluxetable}


\clearpage

\begin{deluxetable}{llcrrrccccr}

\tablecaption{Anderson-Darling Probabilities\tablenotemark{a}
\label{AD}}
\tablewidth{0pt}
\tablehead{ \colhead{Group} &  \colhead{Jsat}  & \colhead{Ssat}  & \colhead{Usat} & \colhead{Nsat} & \colhead{JTro}   }
\startdata
Jsat   	& 1.000  &   0.130   & 0.005 &    0.008  &  0.704 \\
Ssat   	&  & 1.000 & 0.310 & 0.087 & 0.261   \\
Usat  &  &  & 1.000 & 0.319   & 0.389   \\
Nsat   &    &   &   &  1.000  & 0.142  \\
Jtro        &    &   &   &    & 1.000\\

\enddata

\tablenotetext{a}{Probability that any two given color distributions could be drawn from the same parent population.   The lower half of the diagonally symmetric matrix is not shown.}

\end{deluxetable}

\clearpage

\clearpage

\clearpage

\clearpage

\clearpage

\clearpage

\clearpage

\clearpage

\clearpage

\clearpage

\end{document}